\documentclass[twocolumn]{revtex4-1}

\usepackage{amssymb}
\usepackage[utf8]{inputenc}
\usepackage[usenames]{color}
\usepackage{mathptmx}
\usepackage{array}
\usepackage{graphicx}
\usepackage[colorlinks=true] {hyperref}
\usepackage{xspace}
\usepackage{tabularx}
\usepackage{color}
\usepackage[TABBOTCAP]{subfigure}
\usepackage{multirow}
\usepackage{colortbl}

\newcommand{\LMOPS}{Laboratoire Matériaux Optiques, Photonique et Systèmes}
\newcommand{\LMOPSaddress}{Metz, F-57070, France}

\newcommand{\LMOPSUL}{Universit\'e de Lorraine, \LMOPS, \LMOPSaddress}

\newcommand{\LMOPSCS}{\LMOPS, CentraleSupélec, Université Paris-Saclay, \LMOPSaddress}

\newcommand{\E}[2]{#1\times10^{#2}}
\newcommand{\Unit}[2]{ {#1}^{(\mathrm{ #2})}  }

\newcommand{\Atlas}{Atlas$^{\mathrm{\textregistered}}$\xspace}
\newcommand{\Silvaco}{Silvaco$^{\mathrm{\textregistered}}$\xspace}

\newcommand{\X}{{x}}
\newcommand{\Ln}{L_n}

\newcommand{\Li}{L_i}
\newcommand{\WF}{W_f}
\newcommand{\Nd}{N_d}

\newcommand{\Ni}{N_i}

\begin{document}

\title{InGaN Metal-IN Solar Cell: optimized efficiency and fabrication tolerance}

\author{Abdoulwahab Adaine}
\author{Sidi Ould Saad Hamady}
\author{Nicolas Fressengeas}

\affiliation{\LMOPSUL}

\affiliation{\LMOPSCS}

\begin{abstract}

Choosing the Indium Gallium Nitride (InGaN) ternary alloy for thin films solar cells might yield high benefits concerning efficiency and reliability, because its bandgap can be tuned through the Indium composition and radiations have little destructive effect on it. It may also reveal challenges because good quality p-doped InGaN layers are difficult to elaborate. In this letter, a new design for an InGaN thin film solar cell is optimized, where the p-layer of a PIN structure is replaced by a Schottky contact, leading to a Metal-IN (MIN) structure. With a simulated efficiency of $19.8\%$, the MIN structure performs better than the previously studied Schottky structure, while increasing its fabrication tolerance and thus functional reliability\footnote{This journal article is written on the basis of the research results presented during the $\mathrm{2^{nd}}$ Edition Nanotech France 2016 that was held from $\mathrm{1^{st}}$ to $\mathrm{3^{rd}}$ June 2016 in Paris- France}.

\end{abstract}



%

\maketitle


Owing to its good tolerance to radiations \cite{polyakov2013radiation}, its high light absorption \cite{matioli2011high, lin2012simulation} and its Indium--composition--tuned bandgap \cite{bhuiyan2012ingan,reichertz_demonstration_2009}, the Indium Gallium Nitride (InGaN) ternary alloy is a good candidate for high--efficiency--high--reliability solar cells able to operate in harsh environments.

Unfortunately, InGaN p-doping is still a challenge, owing to InGaN residual n-doping \cite{pantha2011origin}, the lack of dedicated acceptors \cite{dahal_ingan_gan_2009} and the complex fabrication process itself \cite{meng2010mg, gherasoiu2014ingan}. To these drawbacks can be added the uneasy fabrication of ohmic contacts \cite{bhuiyan2012ingan} and the difficulty to grow the high-quality-high-Indium-content thin films \cite{yamamoto_metal-organic_2013} which would be needed to cover the whole solar spectrum. These drawbacks still prevent InGaN solar cells to be competitive with other well established III-V and silicon technologies \cite{toledo_ingan_2012}.

In this letter, is proposed a new Metal-IN (MIN) InGaN solar cell structure where the InGaN p-doped layer is removed and replaced by a Schottky contact, lifting one of the above mentioned drawbacks.
A set of realistic physical models based on actual measurements is used to simulate and optimize its behavior and performance using mathematically rigorous multi-criteria optimization methods, aiming to show that both efficiency and fabrication tolerances are better than the previously described simple InGaN Schottky solar cell \cite{ould_saad_hamady2016numerical}.


The material dependent parameters used in this study have been determined for GaN and InN binaries, either from experimental work or \emph{ab initio} calculations \cite{nawaz_tcad_2012,brown_finite_2010}. A review of their values is given in Table \ref{tab_param}. The values for InGaN were linearly interpolated in between the GaN and InN binaries, except for the bandgap $E_g$ and the electronic affinity $\chi$ where the modified Vegard Law was used, with a bowing factor $b=1.43 \mathrm{eV}$ for the bandgap and $b=0.8 \mathrm{eV}$ for the affinity, respectively \cite{franssen2008bowing,brown_finite_2010}. 

In the InGaN III-Nitride semiconductor, the transport equations for electrons and holes can be derived from a drift-diffusion model, provided both carriers mobilities are deduced from temperature and doping using the Caughey-Thomas expressions \cite{schwierz_electron_2005}
\newcommand{\TRatio}{\left(\frac{T}{300}\right)}
\begin{equation}
\label{eq_mobility}
  \mu_m=\mu_{1m}\TRatio^{\alpha_m}+\frac{\mu_{2m}\TRatio^{\beta_m}- \mu_{1m}\TRatio^{\alpha_m}}{1+\left(\frac{N}{N_m^{\mathrm{crit}}\TRatio^{\gamma_m}}\right)^{\delta_m}}
,
\end{equation}
 in which $m$ is either $n$ or $p$, $\mu_n$ being the electrons mobility and $\mu_p$ that of holes. $T$ is the absolute temperature. $N$ is the doping concentration. $N^{\mathrm{crit}}$ and the $n$ or $p$ subscripted $\alpha$, $\beta$, $\delta$ and $\gamma$ are the model parameters which depend on the Indium composition \cite{brown_finite_2010}. Their values have been extracted from the literature, as detailed in tables \ref{tab_mobilparam_n} and \ref{tab_mobilparam_p}.

To increase the carrier transport modeling precision above the mere change in the mobility, were included in the model the bandgap narrowing effect \cite{schenk2008band}, as well as the Shockley–Read–Hall (SRH) \cite{ryu2009rate} and the direct and Auger recombination models using Fermi statistics \cite{bertazzi2010numerical}. To complete the picture, the holes and electrons lifetime was taken equal to 1ns \cite{kumakura_minority_2005} in InGaN.

\begin{table}
  
\begin{center}
\subtable[Data  from refs \cite{brown_finite_2010}.]{
\begin{tabular}{|c|c|c|c|c|c|}
\hline
~ & $\Unit{E_g}{eV}$ & $\Unit{\chi}{eV}$ & $\Unit{N_c}{cm^{-3}}$ & $\Unit{N_v}{cm^{-3}}$ & $\varepsilon$ \\ \hline
GaN & 3.42 & 4.1 & $\E{2.3}{18}$ & $\E{4.6}{19}$ & 8.9 \\ \hline
InN & 0.7 & 5.6 & $\E{9.1}{17}$ & $\E{5.3}{19}$ & 15.3  \\ \hline
    \end{tabular}

\label{tab_physparam}
} 

\subtable[Data from refs \cite{nawaz_tcad_2012,wang2014analytical}.]
{\begin{tabular}{|c|c|c|c|c|c|c|c|}
\hline
~ &  $\Unit{\mu_n^1}{cm^2/Vs}$ &  $\Unit{\mu_n^2}{cm^{2}/Vs}$   & $\delta_n$ & $\Unit{N_n^{\mathrm{crit}}}{cm^{-3}}$ \\ \hline
GaN & 295  & 1460  & 0.71 & $\E{7.7}{16}$\\ \hline
InN & 1982.9  & 10885  & 0.7439 & $\E{1.0}{17}$ \\ \hline
    \end{tabular}
\label{tab_mobilparam_n}}

\subtable[Data from ref \cite{brown_finite_2010}.]
{\begin{tabular}{|c|c|c|c|c|c|c|c|}
\hline
~ &  $\Unit{\mu_p^1}{cm^2/Vs}$ &  $\Unit{\mu_p^2}{cm^{2}/Vs}$   & $\delta_p$ & $\Unit{N_p^{\mathrm{crit}}}{cm^{-3}}$ \\ \hline
GaN & 3.0  & 170  & 2.0 & $\E{1.0}{18}$\\ \hline
InN & 3.0  & 340  & 2.0 & $\E{8.0}{17}$ \\ \hline
    \end{tabular}
\label{tab_mobilparam_p}}
\end{center}
\caption{Experimental or \emph{ab initio} data used in the simulations. Owing to the absence of any experimental data, $\alpha_n$, $\beta_n$, $\gamma_n$, $\alpha_p$, $\beta_p$ and $\gamma_p$ have been estimated to 1. \label{tab_param}}
\end{table}

\begin{table}
  \begin{center}
\begin{tabular}{|l|c|c|}
\hline
Indium Composition & $\Unit{C}{eV^{-1}}$ & $\Unit{D}{eV^{-2}}$ \\ \hline
1	&0.69642	&0.46055\\\hline
0.83	&0.66796	&0.68886\\\hline
0.69	&0.58108	&0.66902\\\hline
0.57	&0.60946	&0.62182\\\hline
0.5	&0.51672	&0.46836\\\hline
0	&3.52517	&-0.65710\\\hline
  \end{tabular}
  \end{center}
\caption{Values for $C$ and $D$ in equation (\ref{eq_absorption}) as found by Brown \emph{et. al.} in \cite{brown_finite_2010}.}
\label{tab_Brown}
\end{table}

Light absorption in InGaN is modeled for the whole solar spectrum and for all $x$ Indium compositions   using a previously proposed phenomenological model \cite{brown_finite_2010} as
\begin{equation}
  \Unit{\alpha}{cm^{-1}}=\Unit{10^5}{cm^{-1}}\sqrt{C\left(E_{ph}-E_g\right)+D\left(E_{ph}-E_g\right)^2},
  \label{eq_absorption}
\end{equation}
where $E_{ph}$ is the incoming photon energy, $E_g$ is the material bandgap at a given Indium composition, $C$ and $D$ are empirical parameters depending on the Indium composition. They are modeled from experimental measurements\cite{brown_finite_2010} summarized in Table \ref{tab_Brown}. Their dependence on the Indium composition $\X$ is approximated by a polynomial fit, of the $\mathrm{4}^{th}$ degree for the former, and quadratic for the latter:
\begin{eqnarray*}
C &=& 3.525 - 18.29\X + 40.22\X^2 - 37.52\X^3 + 12.77\X^4\\
D &=& -0.6651 + 3.616\X - 2.460\X^2
\end{eqnarray*}

The refraction index is modeled through the Adachi model \cite{djurisic_modeling_1999} defined for InGaN and for a given photon energy as
\newcommand{\ERatio}{\frac{E_{ph}}{E_g}}
\begin{equation}
  n\left(E_{ph}\right)=\sqrt{\frac{A}{\left(\ERatio\right)^2}\left[2-\sqrt{1+\ERatio}-\sqrt{1-\ERatio}\right]+B },
  \label{eq_index}
\end{equation}
where $A$ and $B$ are also empirical parameters depending on the Indium composition. They have been experimentally measured \cite{nawaz_tcad_2012,brown_finite_2010} for GaN ($A^{\mathrm{GaN}}=9.31$ and $B^{\mathrm{GaN}}=3.03$) and InN ($A^{\mathrm{InN}}=13.55$ and $B^{\mathrm{InN}}=2.05$) and are linearly interpolated for InGaN

Finally, a ASTM-G75-03 solar spectrum taken from the National Renewable Energy Laboratory database \footnote{\url{http://rredc.nrel.gov/solar/spectra/am1.5/astmg173/astmg173.html}} was shone on the solar cell.

The devices were then simulated in the framework of the above mentioned drift-diffusion model using the \Atlas device simulation tool from \Silvaco, in which the above described detailed physical model was implemented. Solving the coupled drift-diffusion equations in two dimensions allowed the calculation of the solar cell performances, along with its spectral response, I-V characteristics, electric field and potential distributions\dots

The mathematically rigorous L-BFGS-B quasi-Newton optimization method \cite{nocedal2006large} was used to find the optimum efficiency with respect to a given set of parameters; work done through a Python package that we developed in the SAGE \cite{sage} interface to the SciPy \cite{van_der_walt_numpy_2011,scipy} optimizers, using the \Atlas simulator as the backend engine.


\begin{figure}
\includegraphics[width=\linewidth]{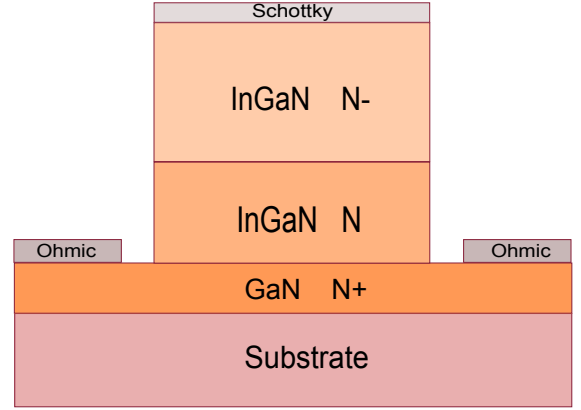}
\label{MIN}
\caption{Schematic view of the MIN solar cell structure.}
\end{figure}

\begin{table*}
	\setlength{\tabcolsep}{3pt}
	\footnotesize
	\centering
	\renewcommand{\arraystretch}{1.0}

	\resizebox{\textwidth}{!}{\begin{tabular}{|>{\bfseries}l|c|c|c|c|c|c|c|}

	\cline{2-8}\multicolumn{1}{c|}{} &  &  &  &  &  &  & $\bm{\eta(\%)}$ \\

	\multicolumn{1}{c|}{} & $\bm{\Li(\mu m)}$ & $\bm{\Ln(\mu m)}$ & $\bm{\Ni(cm^{-3})}$ & $\bm{\Nd(cm^{-3})}$ & $\bm{\WF(eV)}$ & $\bm{x}$ &  $\bm{V_{OC}(V)}$ \\

	\multicolumn{1}{c|}{} &  &  &  &  &  &  & $\bm{J_{SC}(mA/cm^2)}$ \\

	\multicolumn{1}{c|}{} &  &  &  &  &  &  & $\bm{FF(\%)}$ \\

	\hline
		
		Range & $\bm{[0.10 - 1.00]}$ & $\bm{[0.10 - 1.00]}$ & $\bm{[1.0\times10^{14} - 1.0\times10^{17}]}$ & $\bm{[1.0\times10^{16} - 1.0\times10^{19}]}$ & $\bm{[5.50 - 6.30]}$ & \shortstack[c]{\\ $\bm{[0.00 - 1.00]}$} & \\

	\hline
	
	       &  &  &  &  &  &  & $\bm{19.8}$ \\
	
      MIN & $\bm{0.61}$ & $\bm{0.83}$ & $\bm{6.1\times10^{16}}$ & $\bm{3.6\times10^{17}}$ & $\bm{6.30}$ & $\bm{0.60}$ & $\bm{0.835}$ \\

	       & $\bm{[0.10-1.00]}$ & $\bm{[0.10 - 1.00]}$ & $\bm{[1.0\times10^{14}-1.0\times10^{17}]}$ & $\bm{[1.8\times10^{16}-1.0\times10^{19}]}$ & $\bm{[6.11-6.30]}$ & $\bm{[0.48-0.72]}$ & $\bm{30.29}$ \\

	       &  &  &  &  &  &  & $\bm{78.39}$ \\

	\hline
	
	& \cellcolor[gray]{.9} &  & \cellcolor[gray]{.9} &  &  &  & $\bm{18.2}$ \\

 Schottky & \cellcolor[gray]{.9} & $\bm{0.86}$ & \cellcolor[gray]{.9} & $\bm{6.5\times10^{16}}$ & $\bm{6.30}$ & $\bm{0.56}$ & $\bm{0.863}$ \\

	 \cite{ould_saad_hamady2016numerical} & \cellcolor[gray]{.9} & $\bm{[0.53 - 1.00]}$ & \cellcolor[gray]{.9} & $\bm{[1.0\times10^{16}-3.0\times10^{17}]}$ & $\bm{[6.15-6.30]}$ & $\bm{[0.50-0.72]}$ & $\bm{26.80}$ \\

	       & \cellcolor[gray]{.9} &  & \cellcolor[gray]{.9} &  &  &  & $\bm{78.82}$ \\

	\hline

	\end{tabular}}

	\caption{Optimum efficiency $\eta$ obtained for the $MIN$ solar cell and associated open-circuit voltage $V_{OC}$, short-circuit current $J_{SC}$ and Fill Factor $FF$, along with the corresponding physical and material parameters, all compared to the previously published $Schottky$ structure \cite{ould_saad_hamady2016numerical} used as a reference. These results are obtained from several optimizations with random starting points ensuring the absoluteness of the optimum efficiency $\eta$. $\X$ is the indium composition. $\Li$ and $\Ln$ are the thicknesses of the $I$ and $N$ layers respectively and where applicable. $\Ni$ and $\Nd$ are the dopings of the $I$ and $N$ layers respectively where applicable. For each parameter, a range and a tolerance range are given. The range is on the second line of the table. It is the range within which the optimum value of a given parameter is sought. The tolerance range is given just below each parameter optimal value. It corresponds to the set of values of that parameter for which the efficiency $\eta$ remains above $90\%$ of its maximum, the other parameters being kept at their optimum values. }
	\label{tab_optimum}

\end{table*}

\begin{figure}
	\includegraphics[width=\linewidth]{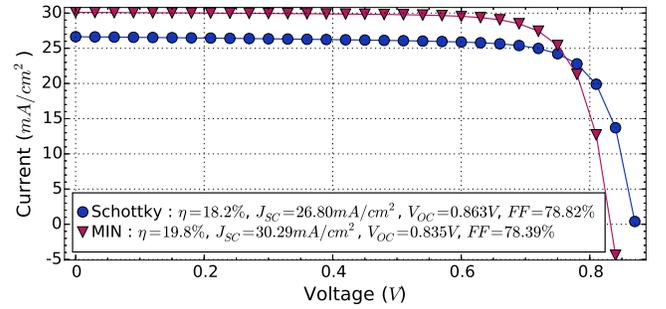}
	\caption{Current-voltage characteristic for the $InGaN$ MIN solar cell compared to the Schottky one.}
\label{MIN_rev_jv}
\end{figure}

Unlike the n-type doping, relatively easy for the InGaN alloy, the p-type doping is still challenging to achieve, owing mainly to the unintentional n-doping (UID) and the lack of adequate acceptors \cite{dahal_ingan_gan_2009}. 
The $MIN$ solar cell was optimized with respect to its most important parameters: $\Li$ and $\Ln$, the thicknesses of the $I$ and $N$ layers respectively, $\Ni$ and $\Nd$, the doping levels of the $I$ and $N$ layers respectively, the Indium composition $\X$ and the metal workfunction $\WF$. The optimal values for all these parameters have been sought within a physically and technologically meaningful interval.
The resulting optimum efficiency is reported in table \ref{tab_optimum}, along with the associated photovoltaic parameters as well as the corresponding parameters and their tolerance range. The results corresponding to the previously reported Schottky structure  \cite{ould_saad_hamady2016numerical} are provided for comparison purposes.
The maximum MIN cell efficiency is found to be $19.8\%$, comparable to  the highest efficiencies reported for the thin films solar cells \cite{green_solar_2015}.

Figure \ref{MIN_rev_jv} shows the current-voltage characteristics of the  optimal $MIN$ solar cell, the Schottky one still being shown for comparison purposes. We observe that the $MIN$ structure has a higher $J_{SC}$ compared to the $Schottky$ structure, but a lower $V_{OC}$, associated to a higher overall efficiency. This behavior stems from the change in the bandgap induced by the Indium composition variation.

\begin{figure}
\includegraphics[width=\linewidth]{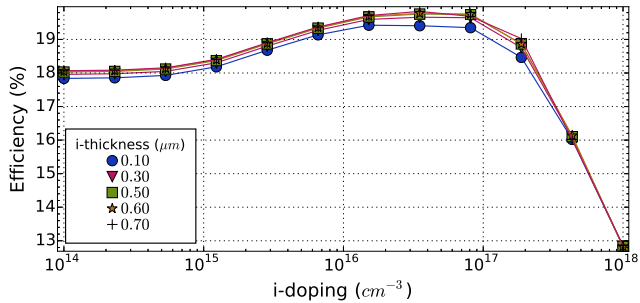}
\caption{InGaN MIN solar cell efficiency for the optimal parameters, varying only the i-layer parameters. Only one third of the points is plotted for clarity's sake.}
\label{MIN_rev_eff}
\end{figure}

Figure \ref{MIN_rev_eff} shows the variation of the photovoltaic (PV) efficiency as a function of the i-layer doping (i-doping) for various i-thicknesses, the other parameters being at their optimal value. The optimal i-doping value is about $6.1\times 10^{16} \mathrm{cm^{-3}}$. However, figure \ref{MIN_rev_eff} shows that choosing lower i-dopings does not impact the efficiency too much. On the contrary, choosing higher dopings quickly and drastically reduces the PV efficiency.

\begin{figure}
\includegraphics[width=\linewidth]{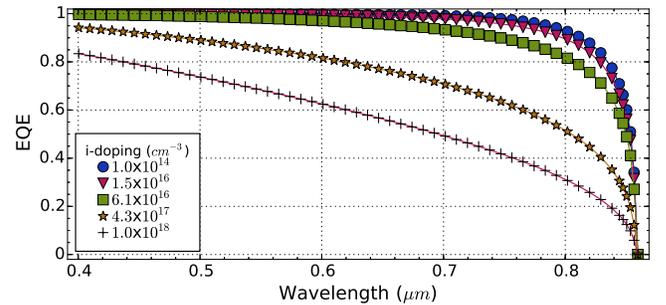}
\caption{InGaN MIN solar cell external quantum efficiency (EQE) spectra for the optimal parameters, varying only the i-layer doping concentration around its optimal $6.1\times 10^{16}\mathrm{cm}^{-3}$ value from a low $1.0\times 10^{14}\mathrm{cm}^{-3}$ to a high $1.0\times 10^{18}\mathrm{cm}^{-3}$ . For clarity's sake, only one calculated point out of three is shown.}
\label{MIN_rev_eqe}
\end{figure}

Figure \ref{MIN_rev_eqe} shows the External Quantum Efficiency (EQE) spectra of the optimal MIN solar cell for an i-layer doping ranging from $1.0\times 10^{14}\mathrm{cm}^{-3}$ to  $1.0\times 10^{18}\mathrm{cm}^{-3}$.
 As can be seen from figure \ref{MIN_rev_eqe},  the optimal doping of $6.1\times 10^{16}\mathrm{cm}^{-3}$ does not yield the optimal EQE. Indeed, an optimum EQE corresponds to an optimal photocurrent, while an increase in the i-doping also implies here a raise in the solar cell voltage. This results in a trade-off between increasing voltage and decreasing photocurrent, yielding an intermediate i-doping optimum. 
At this optimal i-doping, the EQE value is close to its maximum value for a large fraction of the solar spectrum. 

The ultimate goal of this work is the actual device solar cell fabrication. That is the reason why the simulations and optimizations have been conducted with actually measured parameters and realistic physical models. To complete the study on actual fabrication, we have a conducted a tolerance analysis on the optimal parameters that were found. We have thus defined a tolerance range, which is the range of values of a given parameter for which the efficiency $\eta$ remains above $90\%$ of its maximum value. The tolerance range is shown on table \ref{tab_optimum}, just below the optimal value. For instance, for the $MIN$ structure, the efficiency value remains between $17.8\%$ and $19.2\%$ for an i-layer doping $\Ni$ varying between $1.0\times10^{14} \mathrm{cm^{-3}}$ and $1.0\times10^{17} \mathrm{cm^{-3}}$, the other parameters remaining at their optimal values.

The MIN structure tolerance ranges, wider than that of the Schottky structure \cite{ould_saad_hamady2016numerical}, allow to remove another drawback of solar cell InGaN technology, which is the difficult realization of ohmic contacts. Indeed, the wide tolerance range on the n-doping $[1.8\times10^{16}-1.0\times10^{19}] \mathrm{cm^{-3}}$ allows to design heavily doped n-layers to elaborate low resistance ohmic contacts on InGaN without noticeably impacting the photovoltaic performances.

Finally, and as hinted previously, we attempt to address another of the main challenges preventing the development of high efficiency InGaN solar cell, which is the high defect density usually present in the grown thin films \cite{yamamoto_metal-organic_2013}. In order to study the impact these defects may have on the here proposed MIN solar cell performances, we included in the simulation the dominating deep defects, which have been experimentally studied in literature using the well known Deep Level (Transient \& Optical) Spectroscopy (DLTS and DLOS), the Steady-State PhotoCapacitance (SSPC) and the Lighted Capacitance-Voltage (LCV) techniques \cite{nakano2014electrical, lozac2012study, armstrong2012quantitative, gur2011detailed}. The experimental results resulting from these works are briefly summarized in table \ref{tab_Defect} for the studied Indium compositions.

In order to model these defects for the Indium composition obtained in this work ($0.60$) and the corresponding bandgap ($1.27 \mathrm{eV}$), we reasonably extrapolated the experimental defect energy measured for composition up to $0.20$ and therefore set the defect energy to $0.8 \mathrm{eV}$ below the conduction band edge in the i-layer. To account for probable statistical variations, we used a Gaussian distribution centered at $0.8 \mathrm{eV}$, with a characteristic decay energy $\delta$ varying between $0.01 \mathrm{eV}$ and $0.20 \mathrm{eV}$ and a capture cross section $\sigma$ of $1.1\times10^{-13} \mathrm{cm^{2}}$, which is the highest experimental value reported in \cite{gur2011detailed}. We then evaluated the MIN cell efficiency, varying the total density of states from $1.0\times10^{13} \mathrm{cm^{-3}}$ to $1.0\times10^{17} \mathrm{cm^{-3}}$. This latter density is even higher than the dominating defects concentration reported in \cite{nakano2014electrical, lozac2012study, armstrong2012quantitative, gur2011detailed}.

\begin{table}
	\setlength{\tabcolsep}{3pt}
	\footnotesize
	\centering
	\renewcommand{\arraystretch}{1.0}
\begin{tabular}{|l|c|c|}
\hline
Indium Composition $x$ & Defect energy ($eV$) & Concentration ($\mathrm{cm^{-3}}$) \\ \hline
$0.09$ & $3.05$ & $2.7\times10^{16}$ \\ \hline
$0.13$ & $2.76$ & $8.5\times10^{15}$ \\ \hline
$0.20$ & $2.50$ & $6.1\times10^{16}$ \\ \hline
\end{tabular}
\caption{The dominating deep-level defect parameters in InGaN as experimentally measured and reported in \cite{nakano2014electrical, lozac2012study} for Indium composition $x = 0.09$, in \cite{armstrong2012quantitative} for $x = 0.13$ and in \cite{gur2011detailed} for $x = 0.20$. The defect energy is measured relative to the conduction band edge.}
\label{tab_Defect}
\end{table}

\begin{figure}
\includegraphics[width=\linewidth]{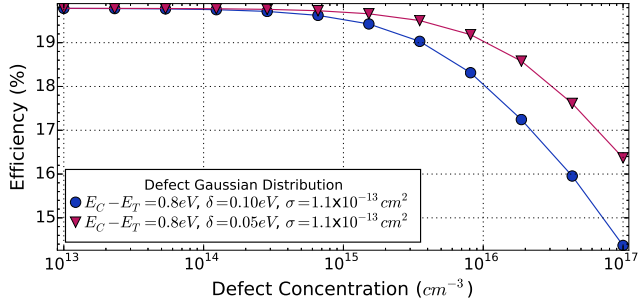}
\caption{InGaN MIN solar cell photovoltaic efficiency variation with defect density for two Gaussian distributions. The cell parameters are fixed to their optimal values shown in table \ref{tab_optimum}.}
\label{MIN_rev_defect}
\end{figure}

Figure \ref{MIN_rev_defect} shows the MIN solar cell photovoltaic efficiency with respect to the defect concentration for two decay energy $\delta$ values of $0.05 \mathrm{eV}$ and $0.10 \mathrm{eV}$. We observe that the solar cell efficiency remains close to its maximum value as long as the defect concentration is smaller than the i-layer doping concentration ($6.1\times 10^{16}\mathrm{cm}^{-3}$). When the defect concentration becomes comparable to the optimal i-layer doping concentration, the solar cell efficiency decreases within a concentration range that depends on the distribution decay energy. This result shows that the defects concentration must be kept lower but not necessarily much lower than the doping concentration. The demonstrated wide tolerance of the MIN structure can allow keeping the defect effect on the overall solar cell efficiency as low as possible by varying accordingly the InGaN doping. A compromise can therefore be found to limit the effect of the defects density that is relatively high in the presently elaborated InGaN layers.


We have thus optimized a new MIN solar cell structure using a rigorous numerical optimization approach and most realistic parameters and physical models. An optimal efficiency of $19.8\%$ was found, associated to wide tolerance ranges, lifting two of the major drawbacks of InGaN technology for solar cells. Indeed, on the one hand, the Schottky contact has removed the need for p-doping, yielding a  MIN solar cell with an efficiency comparable to that of the highest efficiencies reported for the thin films structures. On the other hand, the MIN structure wide tolerances have facilitated the design of low resistance ohmic contacts and the growth defects management.


\bibliographystyle{iopart-num} 
\bibliography{MIN_rev_ref}

\providecommand{\newblock}{}
\begin{thebibliography}{10}
\expandafter\ifx\csname url\endcsname\relax
  \def\url#1{{\tt #1}}\fi
\expandafter\ifx\csname urlprefix\endcsname\relax\def\urlprefix{URL }\fi
\providecommand{\eprint}[2][]{\url{#2}}

\bibitem{polyakov2013radiation}
Polyakov A~Y, Pearton S, Frenzer P, Ren F, Liu L and Kim J 2013 {\em Journal of
  Materials Chemistry C\/} {\bf 1} 877--887

\bibitem{matioli2011high}
Matioli E, Neufeld C, Iza M, Cruz S~C, Al-Heji A~A, Chen X, Farrell R~M, Keller
  S, DenBaars S, Mishra U {\em et~al.\/} 2011 {\em Applied Physics Letters\/}
  {\bf 98} 021102

\bibitem{lin2012simulation}
Lin S, Zeng S, Cai X, Zhang J, Wu S, Sun L and Zhang B 2012 {\em Journal of
  Materials Science\/} {\bf 47} 4595--4603

\bibitem{bhuiyan2012ingan}
Bhuiyan A, Sugita K, Hashimoto A and Yamamoto A 2012 {\em Photovoltaics, IEEE
  Journal of\/} {\bf 2} 276--293

\bibitem{reichertz_demonstration_2009}
Reichertz L~A, Gherasoiu I, Yu K~M, Kao V~M, Walukiewicz W and Ager~III J~W
  2009 {\em Applied Physics Express\/} {\bf 2} 122202

\bibitem{pantha2011origin}
Pantha B, Wang H, Khan N, Lin J and Jiang H 2011 {\em Physical Review B\/} {\bf
  84} 075327

\bibitem{dahal_ingan_gan_2009}
Dahal R, Pantha B, Li J, Lin J and Jiang H 2009 {\em Applied Physics Letters\/}
  {\bf 94} 063505

\bibitem{meng2010mg}
Meng Z, Bhattacharya P, Wei G and Banerjee A 2010 {\em Applied Physics
  Letters\/} {\bf 96}

\bibitem{gherasoiu2014ingan}
Gherasoiu I, Yu K~M, Reichertz L~A and Walukiewicz W 2014 {\em Physica Status
  Solidi (c)\/} {\bf 11} 381--384

\bibitem{yamamoto_metal-organic_2013}
Yamamoto A, Sugita K, Bhuiyan A, Hashimoto A and Narita N 2013 {\em Materials
  for Renewable and Sustainable Energy\/} {\bf 2} 1--9

\bibitem{toledo_ingan_2012}
Toledo N~G and Mishra U~K 2012 {\em Journal of Applied Physics\/} {\bf 111}
  114505

\bibitem{ould_saad_hamady2016numerical}
Ould Saad~Hamady S, Adaine A and Fressengeas N 2016 {\em Materials Science in
  Semiconductor Processing\/} {\bf 41} 219--225

\bibitem{nawaz_tcad_2012}
Nawaz M and Ahmad A 2012 {\em Semiconductor Science and Technology\/} {\bf 27}
  035019

\bibitem{brown_finite_2010}
Brown G, Ager~III J, Walukiewicz W and Wu J 2010 {\em Solar Energy Materials
  and Solar Cells\/} {\bf 94} 478--483

\bibitem{franssen2008bowing}
Franssen G, Suski T, Kami{\'n}ska A, Pereiro~Viterbo J, Mu{\~n}oz~Merino E,
  Lliopoulus E, Georgakilas A, Che S, Ishitani Y, Yoshikawa A {\em et~al.\/}
  2008 {\em Journal of Applied Physics\/} {\bf 103} 033514--1

\bibitem{schwierz_electron_2005}
Schwierz F 2005 {\em Solid State Electronics\/} {\bf 49} 889--895

\bibitem{schenk2008band}
Schenk H, Borenstain S, Berezin A, Sch{\"o}n A, Cheifetz E, Khatsevich S and
  Rich D 2008 {\em Journal of Applied Physics\/} {\bf 103} 103502

\bibitem{ryu2009rate}
Ryu H~Y, Kim H~S and Shim J~I 2009 {\em Applied Physics Letters\/} {\bf 95}
  081114

\bibitem{bertazzi2010numerical}
Bertazzi F, Goano M and Bellotti E 2010 {\em Applied Physics Letters\/} {\bf
  97} 231118

\bibitem{kumakura_minority_2005}
Kumakura K, Makimoto T, Kobayashi N, Hashizume T, Fukui T and Hasegawa H 2005
  {\em Applied Physics Letters\/} {\bf 86} 052105 ISSN 0003-6951, 1077-3118

\bibitem{wang2014analytical}
Wang S, Liu H, Song X, Guo Y and Yang Z 2014 {\em Applied Physics A\/} {\bf
  114} 1113--1117

\bibitem{djurisic_modeling_1999}
Djurišić A~B and Li E~H 1999 {\em Journal of Applied Physics\/} {\bf 85}
  2848--2853

\bibitem{Note1}
\protect \url
  {http://rredc.nrel.gov/solar/spectra/am1.5/astmg173/astmg173.html}

\bibitem{nocedal2006large}
Nocedal J and Wright S~J 2006 {\em Numerical Optimization\/}  164--192

\bibitem{sage}
Stein W {\em et~al.\/} 2014 {\em {S}age {M}athematics {S}oftware ({V}ersion
  6.4.1)\/} The Sage Development Team {\tt http://www.sagemath.org}

\bibitem{van_der_walt_numpy_2011}
van~der Walt S, Colbert S and Varoquaux G 2011 {\em Computing in Science
  Engineering\/} {\bf 13} 22--30 ISSN 1521-9615

\bibitem{scipy}
Jones E, Oliphant T, Peterson P {\em et~al.\/} 2001-- {SciPy}: Open source
  scientific tools for {Python} {\tt http://www.scipy.org/}

\bibitem{green_solar_2015}
Green M~A, Emery K, Hishikawa Y, Warta W and Dunlop E~D 2015 {\em Progress in
  Photovoltaics: Research and Applications\/} {\bf 23} 1--9 ISSN 1099-159X

\bibitem{nakano2014electrical}
Nakano Y, Sang L and Sumiya M 2014 Electrical characterization of thick ingan
  films for photovoltaic applications {\em MRS Proceedings\/} vol 1635
  (Cambridge Univ Press) pp mrsf13--1635

\bibitem{lozac2012study}
Lozac'h M, Nakano Y, Sang L, Sakoda K and Sumiya M 2012 {\em Japanese Journal
  of Applied Physics\/} {\bf 51} 121001

\bibitem{armstrong2012quantitative}
Armstrong A, Henry T~A, Koleske D~D, Crawford M~H and Lee S~R 2012 {\em Optics
  express\/} {\bf 20} A812--A821

\bibitem{gur2011detailed}
G{\"u}r E, Zhang Z, Krishnamoorthy S, Rajan S and Ringel S 2011 {\em Applied
  Physics Letters\/} {\bf 99} 092109

\end{thebibliography}

\end{document}